\documentclass[
prb,twocolumn,preprintnumbers,amsmath,amssymb]{revtex4}

\usepackage{graphicx}
\usepackage{amssymb}
\usepackage{amsmath}
\usepackage{dsfont}
\usepackage{bm}
\usepackage{mathrsfs}
\usepackage{times}

\usepackage{feynmf}
\usepackage{amsfonts}

\begin{document}
\title{
Electromagnetic reflection, transmission and energy density at boundaries of nonlocal media
}

\author{R.\ J.\ Churchill}

\email{rc313@exeter.ac.uk}

\author{T.\ G.\ Philbin}

\email{t.g.philbin@exeter.ac.uk}

\affiliation{Physics and Astronomy Department, University of Exeter,
Stocker Road, Exeter EX4 4QL, United Kingdom}

\begin{abstract}
We consider a semi-infinite spatially dispersive dielectric with unequal transverse and longitudinal susceptibilities. The effect of the boundary is characterized by arbitrary reflection coefficients for polarization waves in the material that propagate to the surface. Specific values of these coefficients correspond to various additional boundary conditions (ABC) for Maxwell's equations. We derive the electromagnetic reflection and transmission coefficients at the boundary and investigate their dependence on material parameters and ABC. We also investigate the electromagnetic zero-point and thermal spectral energy density outside the dielectric. The nonlocal response removes the boundary divergence of the spectral energy density that is present in a local model. The spectral energy density shows a large dependence on the difference between the transverse and longitudinal susceptibilities, even at distances up to 10nm from the boundary.
\end{abstract}

\maketitle

\section{Introduction}\label{sec:introduction}

The response of electromagnetic materials to applied fields is spatially nonlocal, i.e.\ the response at any point depends on the value of the fields throughout a neighboring region\cite{LLcm,rukhadze1961}. This nonlocal response, or spatial dispersion, can be neglected in many regimes of interest, even when frequency dispersion is significant. Nevertheless, there are cases where nonlocal response is a significant factor and interest in this topic has increased in recent years. Metallic nano-structures have been developed whose properties can be accurately predicted only by including their nonlocal  response\cite{raz11,wie12,pen12,tos15,sch16}. Near-field radiative heat transfer between materials is also modified by spatial dispersion\cite{sin15a,sin15b}. More generally, thermal and zero-point electromagnetic energy in the presence of materials is significantly affected by spatial dispersion\cite{henkel2006,nar10,hor14}. This has implications for spontaneous emission rates of emitters inside materials or placed close to surfaces\cite{pur46,dre68,bar98,Novotny,hor14}, and also for thermal and zero-point forces on curved boundaries\cite{hor14}. There are thus many interesting questions, some of them quite basic, that require a proper account of spatial dispersion.

In this paper we employ the macroscopic Maxwell equations  to explore boundary effects in nonlocal media. We extend previous results on reflection and transmission at planar boundaries to the most general isotropic spatially dispersive dielectric. The key extra ingredient here is to allow for both transverse and longitudinal susceptibilities that have different values\cite{LLcm,rukhadze1961} (see below). We also show how the spectral energy density of thermal and zero-point radiation depends on the material susceptibilities in this  general case.

The electric susceptibility $\chi$ is frequently described by a damped-oscillator model, in which nonlocal response may be incorporated by a simple wave-vector dependence\cite{hopfieldthomas} :
\begin{align}
\chi({\bm k},\omega)
=
\chi_0
+
\frac{\omega_p^2}{\omega_T^2+\sigma^2k^2-\omega^2-i\gamma\omega}.
\label{eq:susceptibility_basic}
\end{align}
Here $\omega_T$ is the resonant frequency, $\gamma$ quantifies the absorption, $\omega_p$ is the oscillator strength, and $\sigma$ is a spatial-dispersion parameter. The term $\chi_0$ collects contributions from other resonances and acts as a background susceptibility. A more complete model would dispense with  $\chi_0$ and include additional resonance terms.  A justification of (\ref{eq:susceptibility_basic}) based on properties of semi-conductors was given by Hopfield and Thomas~\cite{hopfieldthomas} but it can also be derived from a simple classical model~\cite{chu16}. In the local case $\sigma=0$ the usual Maxwell boundary conditions, namely the continuity of the tangential components of ${\bm E}$ and ${\bm H}$ and the normal components of ${\bm D}$ and ${\bm B}$, are sufficient to calculate the reflection coefficient at a sharp boundary. However, the introduction of a nonlocal term as in (\ref{eq:susceptibility_basic}) leads to the presence of two transverse and one longitudinal wave inside the medium\cite{rukhadze1961}.
The usual Maxwell boundary conditions are now insufficient to solve for the four unknown amplitudes of the transmitted and reflected waves. Additional information is required about the relationship between the amplitudes. Many authors have expressed this as additional boundary conditions (ABC's) on the polarization ${\bm P}$ of the medium associated with the spatially dispersive resonance\cite{agarwal1971a,agarwal1971b,agarwal1972,birman1972,agarwal1973,maradudin1973,birman1974,mills1974,foley1975,bishop1976,ting1975,kliewer1968,kliewer1971,kliewer1975,ruppin1981,rimbey1974,rimbey1975,rimbey1976,rimbey1977,rimbey1978,pekar1958a,pekar1958b,pekar1958c,pekar1959}
 (i.e.\ without the background term $\chi_0$).
Each ABC is motivated by the type of medium considered, as we describe in more detail in Sec. \ref{sec:p-polarization}. Several of these authors note that ABCs are equivalent to introducing a phenomenological scattering term to the susceptibility in the presence of a boundary, in order to describe the behavior of the medium at the surface. The denominator of the second term in (\ref{eq:susceptibility_basic}) has zeros that correspond to the dispersion relation for waves of polarization $\bm{P}$ in the material. These polarization waves are reflected at the surface and complex parameters are introduced to serve as the corresponding reflection coefficients. Halevi and Fuchs\cite{Halevi} incorporated all previous examples of these extra parameters into values of a set $U_i$ ($i\in\{x,y,z\}$)  of reflection coefficients for the polarization waves (see below). They then derived a general expression for the electromagnetic reflection coefficients at the boundary in terms of arbitrary complex $U_i$, in the case of a dielectric whose bulk susceptibility is the scalar $\chi({\bm k},\omega)$ of the form (\ref{eq:susceptibility_basic}).

However, in the presence of spatial dispersion the susceptibility is a tensor, as the wave-vector generates a distinctive direction\cite{LLcm}. In a homogeneous, isotropic, non-gyroscopic medium:
\begin{align}
\chi_{ij}({\bm k},\omega)
=
\delta_{ij}
\chi_{\perp}({\bm k},\omega)
+
\frac{k_ik_j}{k^2}
\left[
\chi_{\parallel}({\bm k},\omega)
-
\chi_{\perp}({\bm k},\omega)
\right],
\label{eq:susc}
\end{align}
where $\perp$ and $\parallel$ denote transverse and longitudinal terms, respectively. If one assumes $\chi_{\parallel}=\chi_{\perp}$ then the susceptibility (\ref{eq:susc}) is essentially still a scalar, but the most general isotropic susceptibility is a tensor in the nonlocal case.
In comparison to the simplified scalar case, the tensor nature of the susceptibility is generally overlooked.
While Rimbey and Mahan include this in their calculation for a specific ABC\cite{rimbey1974}, their choice of $U_i$ leads to the absence of a longitudinal wave in the medium.
Garcia-Moliner and Flores\cite{garcia1977} derive the reflection coefficient in the tensor case, but they restrict themselves to a scalar $U$ and obtain a result in an integral form.

The first aim of this paper is to extend Halevi and Fuchs' derivation\cite{Halevi} to the tensor susceptibility (\ref{eq:susc}), with $\chi_{\parallel}\neq\chi_{\perp}$, where $\chi_{\parallel}$ and $\chi_{\perp}$ each have the form (\ref{eq:susceptibility_basic}). We derive a general expression for reflection and transmission coefficients at a planar boundary, allowing for arbitrary complex reflection coefficients of the polarization waves at the surface. 

Our second goal is to use the general electromagnetic reflection coefficients derived in the first part of the paper to calculate the spectral energy density of  thermal and zero-point radiation outside the boundary of the spatially dispersive medium. It is well known that the result for a local medium is proportional to $1/z^3$ close to the surface\cite{candelas1982,henkel2000,joulain2005}. Note that the divergence of the spectral energy on the boundary also occurs for purely thermal radiation (dropping the zero-point part) so it is not due to the divergence of (total) vacuum energy. This unphysical divergence can be removed by introducing a cutoff wave-vector \cite{candelas1982} based on the interatomic or lattice spacing. But a more accurate picture is obtained by the inclusion of spatial dispersion, which must naturally remove the divergence without the need for additional modifications to the calculation. This has been shown\cite{henkel2006} to work  for the Lindhard susceptibility of a plasma, which can be used to model the response of the conduction electrons in a metal. The susceptibility due to the core electrons, however, will still lead to a divergence if it is taken to be local. We will show that the general reflection coefficients derived here give a finite thermal and zero-point spectral energy density at a planar boundary. Moreover, we find that the difference  $\chi_{\parallel}-\chi_{\perp}$ between the transverse and longitudinal susceptibilities has a large effect on the spectral energy close to the boundary. The influence of a metal boundary on the spectral thermal energy density has been measured using near-field microscopy\cite{wilde06}. 

As our treatment is based on macroscopic electromagnetism, we do not include quantum mechanical features, such as ``electron spill-out''. that are not directly encoded in the bulk susceptibility. This means that our results for quantities close to a sharp boundary will lose accuracy below a few nanometers. In practice however, it has been found that some quantum features of the surface can be incorporated through a spatially dispersive susceptibility\cite{wie12}. Our work also assumes a smooth boundary, but surface roughness can potentially be incorporated in a similar fashion to that employed for a local medium\cite{halevi93,biehs11}. Boundary layers containing slits\cite{ren15} or other nontrivial structures\cite{xiong13} would require additional considerations of the field behavior in the interface layer. Our model can be used to find the reflection and transmission coefficients for spatially dispersive metamaterials when the wavelength is such that an effective medium description can be used. Finally, for materials such as thin films or nanospheres, a different approach to that used here is required because of more complicated possibilities for the behavior of polarization waves (e.g.\ multiple reflections from closely separated boundaries).

The paper is organized as follows.
In Sec. \ref{sec:model} we present the spatially dispersive susceptibility model for a half-infinite dielectric with a tensor permittivity and derive the field equations.
In Secs. \ref{sec:p-polarization} and \ref{sec:transmission} we derive the general expressions for the reflection and transmission coefficients and present the results for a variety of ABC's. In Sec. \ref{sec:energy} we calculate the zero-point and thermal spectral energy density and show in detail how the nonlocal response removes the divergence in this quantity that is present in a local model.

\section{Dielectric Model}\label{sec:model}

We first consider an infinite, homogeneous, spatially-dispersive dielectric with the susceptibility (\ref{eq:susc}).
The electric field ${\bm E}$ and polarization field ${\bm P}$ satisfy the wave equation:
\begin{align}
{\bm \nabla}
\times
{\bm \nabla}
\times
{\bm E}({\bm r},\omega)
-
\frac{\omega^2}{c^2}
{\bm E}({\bm r},\omega)
=
\frac{\omega^2}{c^2}
{\bm P}({\bm r},\omega),
\label{eq:wave_equation}
\end{align}
where the polarization field is:
\begin{align}
P_i({\bm r},\omega)=
\int  d^3{\bm r}^\prime
\sum_j
\chi_{ij}({\bm r}-{\bm r}^\prime,\omega)
E_j({\bm r^\prime},\omega).
\label{eq:infinite_polarization}
\end{align}
Using the Fourier transformation:
\begin{align}
P_i({\bm r},\omega)
=
\frac{1}{(2\pi)^3}
\int
d^3{\bm k}
P_i({\bm k},\omega)
e^{i {\bm k} \cdot {\bm r}}
\end{align}
we have:
\begin{align}
P_i({\bm k},\omega)
=
\sum_j
\chi_{ij}({\bm k},\omega)
E_j({\bm k},\omega).
\end{align}
The wave equation (\ref{eq:wave_equation}) has solutions for  ${\bm E}$ when the frequency and wave vector satisfy the dispersion relation\cite{rukhadze1961}:
\begin{align}
(\omega/c)^2\left[1+\chi_\perp({\bm k},\omega)\right]=k^2,
\label{eq:transverse_disp_rel}
\end{align}
for transverse waves with ${\bm E}\cdot{\bm k}=0$ or:
\begin{align}
1+\chi_\parallel({\bm k},\omega)=0,
\label{eq:longitudinal_disp_rel}
\end{align}
for longitudinal waves with ${\bm E}\times{\bm k}=0$.
As the electric field is parallel to the wave vector for the longitudinal wave, this wave has no magnetic field.
With an $\textrm{exp}(ik_zz)$ field dependence we restrict ourselves to wave vectors with $\textrm{Im}[k_z]>0$. There are two solutions to (\ref{eq:transverse_disp_rel}) which we denote ${\bm k_1}$, ${\bm k_2}$ and one solution to (\ref{eq:longitudinal_disp_rel}) which we denote ${\bm k_3}$.

\begin{figure}[!htb]\centering
{\includegraphics[width=\linewidth]{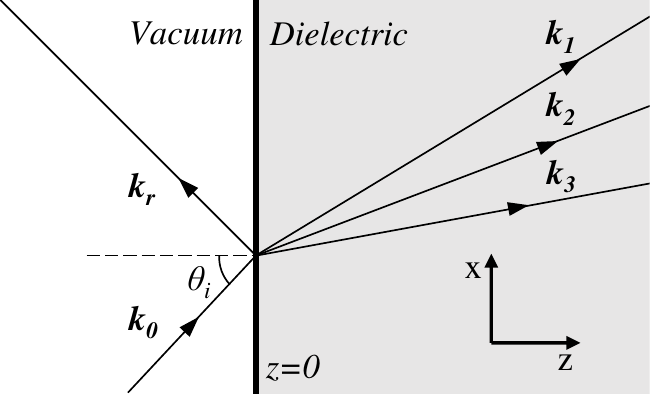}}
\caption{
Schematic of the model.
The $z<0$ vacuum half-space contains the incident (${\bm k_0}$) and reflected (${\bm k_r}$) wave.
The angle of incidence is $\theta_i$.
The $z>0$ spatially dispersive dielectric half-space contains two transverse (${\bm k_1}$, ${\bm k_2}$) and one longitudinal (${\bm k_3}$) transmitted waves.
The coordinate system is chosen such that the $xz$-plane coincides with the plane of incidence and $k_y=0$.
}
\label{fig:model}
\end{figure}

We now consider a half-infinite dielectric that occupies the $z>0$ region as shown in Fig. \ref{fig:model}.
In the vacuum region we have the incident (${\bm E_0}$) and reflected wave (${\bm E_r}$) with wave-vectors $\bm{k}_0$ and $\bm{k}_r$ ($k_0=k_r=\omega/c$), while in the dielectric we have the one longitudinal wave (${\bm E^{(3)}}$) and two transverse waves (${\bm E^{(1)}},{\bm E^{(2)}}$) previously derived.
We choose our coordinate system such that the $xz$-plane coincides with the plane of incidence, with $k_y=0$, $k_x=K$.
The various wave-vectors differ only in the value of $k_z$ .
The angle of incidence is given by $\cos{\theta_i}=\sqrt{k_0^2-K^2}/k_0$. The usual Maxwell boundary conditions are now insufficient to solve for the unknown amplitudes of the three transmitted and one reflected wave.
Additional relationships between the amplitudes are required.
These are usually expressed as additional boundary conditions on the polarization, denoted $\mathcal{P}_i$,  at $z=0^+$ due to the second term in the susceptibility (\ref{eq:susceptibility_basic}):
\begin{align}
\alpha_j\mathcal{P}_j(0^+)
+\beta_j
\partial_z\mathcal{P}_j(0^+)
=0,
\label{eq:robinABC}
\end{align}
for some parameters $\alpha_j$ and $\beta_j$.
A generalized approach was developed by Halevi and Fuchs\cite{Halevi} for a scalar susceptibility, equivalent to (\ref{eq:susc}) with $\chi_\perp=\chi_\parallel$. 
We will now modify their derivation to the tensor case $\chi_{ij}$.

Due to the presence of the boundary, the polarization field  now depends on a position-dependent susceptibility $\chi^\prime$:
\begin{align}
P_i({\bm r})=
\int  d^3{\bm r}^\prime
\sum_j
\chi^\prime_{ij}({\bm r},{\bm r}^\prime)
E_j({\bm r^\prime}),
\end{align}
where we have omitted the $\omega$ dependence for notational simplicity.
After a Fourier transform in the $xy$-plane:
\begin{align}
\tilde{P}_i(k_x,k_y,z)=
\int_{0}^{\infty} dz^\prime
\sum_j
\tilde{\chi}^\prime_{ij}(k_x,k_y,z,z^\prime)
\tilde{E}_j(k_x,k_y,z^\prime).
\label{eq:polarization_half_infinite}
\end{align}
We assume that the overall susceptibility $\chi^\prime$ of the half-infinite dielectric can be expressed in terms of the bulk susceptibility:
\begin{equation}
\tilde{\chi}^\prime_{ij}(k_x,k_y,z,z^\prime)=
\left\{
\begin{aligned}
&\tilde{\chi}_{ij}(k_x,k_y,z-z^\prime)+\\
&U_{ij}\tilde{\chi}_{ij}(k_x,k_y,z+z^\prime),&\text{if } z,z^\prime > 0 \\
&0&\text{otherwise},
\end{aligned}
\right.
\label{eq:susc_U}
\end{equation}
where we have Fourier transformed the bulk susceptibility in (\ref{eq:susc}) to real space in the $z$-direction.
The first term in (\ref{eq:susc_U}) for $z,z^\prime > 0$ is position independent and gives the non-local bulk response.
The second term depends on the distance from the boundary and describes a polarization wave propagating from $z^\prime$ to the surface, reflecting with a (complex in general) amplitude coefficient $U_{ij}$ and continuing to $z$, with $|U_{ij}|=1$ implying elastic reflection. A similar expression to (\ref{eq:susc_U}) had been used previously\cite{garcia1977}, with a scalar $U$ used as a phenomenological description of the dielectric surface response.
Halevi and Fuchs\cite{Halevi} considered a scalar $\chi$, leading to a vector $U_i$ with general values, and showed this to be equivalent to using the ABC's in  (\ref{eq:robinABC}).

After substituting the half-infinite susceptibility (\ref{eq:susc_U}) into (\ref{eq:polarization_half_infinite}), the polarization field takes the form:
\begin{align}
\tilde{P}_i(z)=&
\frac{1}{2\pi}
\int_{-\infty}^{\infty} dq
\int_{0}^{\infty} dz^\prime
\bigg[
e^{iq(z-z^\prime)}
\sum_j
\chi_{ij}(q)
\tilde{E}_j(z^\prime)
\nonumber\\&
+
e^{iq(z+z^\prime)}
\sum_j
U_{ij}
\chi_{ij}(q)
\tilde{E}_j(z^\prime)
\bigg],
\quad
z>0,
\label{eq:polarization_integral}
\end{align}
where $q=k_z$ and we have omitted the $k_x$ and $k_y$ values as they are the same in all arguments.
We now substitute the tensor (\ref{eq:susc}) with expressions for $\chi_\perp$ and $\chi_\parallel$ of the form (\ref{eq:susceptibility_basic}).
The susceptibility (\ref{eq:susceptibility_basic}) can be rewritten as:
\begin{align}
\chi(k_x,k_y,q,\omega)
=
\chi_0
+
\frac{\omega_p^2/\sigma^2}{q^2-\Gamma^2},
\label{eq:susceptibility_simplified}
\end{align}
\begin{align}
\Gamma^2
=
\frac{\omega^2-\omega_T^2+i\gamma\omega-\sigma^2(k_x^2+k_y^2)}{\sigma^2}.
\end{align}
The transverse (longitudinal) susceptibility  takes the form in (\ref{eq:susceptibility_simplified}), but with $\sigma$ and $\Gamma$ replaced by $\sigma_\perp$ ($\sigma_\parallel$) and the corresponding $\Gamma_\perp$ ( $\Gamma_\parallel$).
We define the relationship $\sigma_\parallel^2=(1+\delta)\sigma_\perp^2$, so that the susceptibility tensor reduces to $\delta_{ij}\chi_\perp(k,\omega)$ in the $\delta\to0$ limit.

At this point we introduce an ansatz for the ${\bm E}$ field inside the medium - a linear combination of three plane waves\cite{Halevi}:
\begin{align}
\tilde{E}_j(z)
=
\sum_{n=1}^3
\tilde{E}_j^{(n)}
e^{iq_nz},
\label{eq:E_ansatz}
\end{align}
where $n=1,2$ are the transverse waves and $n=3$ is the longitudinal wave.
Substituting (\ref{eq:E_ansatz}) into (\ref{eq:polarization_integral}) and evaluating the integrals gives:
\begin{align}
\tilde{P}_i(z)=&
\sum_n
\sum_j
\chi_{ij}(q_n)
\tilde{E}^{(n)}_j
e^{iq_nz}
\nonumber\\&
+
\sum_n
\sum_j
\phi_{ij}^{(n)}
\tilde{E}^{(n)}_j
e^{i\Gamma_\perp z}
\nonumber\\&
+
\sum_n
\sum_j
\psi_{ij}^{(n)}
\tilde{E}^{(n)}_j
e^{i\Gamma_\parallel z},
\end{align}
where:
\begin{align}
\phi_{ij}^{(n)}
=&
-
\left[q_n(1+U_{ij})+\Gamma_\perp(1-U_{ij})\right]\chi_\perp(q_n)
\nonumber\\&\times
\frac{1}{2\Gamma_\perp}
\left(\delta_{ij}-\frac{k^{(\perp)}_ik^{(\perp)}_j}{\Gamma_\perp^2+K^2}\right),
\end{align}
\begin{align}
\psi_{ij}^{(n)}
=&
-
\left[q_n(1+U_{ij})+\Gamma_\parallel(1-U_{ij})\right]\chi_\parallel(q_n)
\nonumber\\&\times
\frac{1}{2\Gamma_\parallel}
\left(\frac{k^{(\parallel)}_ik^{(\parallel)}_j}{\Gamma_\parallel^2+K^2}\right),
\end{align}
and ${\bm k}^{(\perp/\parallel)}=\left(K,0,\Gamma_{\perp/\parallel}\right)$.
This must be substituted into the right-hand side (RHS) of the wave equation (\ref{eq:wave_equation}).
All left-hand side terms are proportional to $\textrm{exp}(iq_nz)$, so
for the wave equation to hold for all $z$ values, we require the two RHS sums proportional to $\textrm{exp}(i\Gamma_\perp z)$ and $\textrm{exp}(i\Gamma_\parallel z)$ to equal zero:
\begin{align}
\sum_n
\sum_j
\phi_{ij}^{(n)}
\tilde{E}^{(n)}_j
=
0,
\qquad
\sum_n
\sum_j
\psi_{ij}^{(n)}
\tilde{E}^{(n)}_j
=
0.
\label{eq:abc}
\end{align}
These two equations act as the additional boundary conditions for the system, once the $U_{ij}$ in (\ref{eq:susc_U}) are specified. 

\section{$p$-Polarization}\label{sec:p-polarization}

The wave can be decomposed to components with $\bm{E}$ perpendicular to ($s$-polarized) or in ($p$-polarized) the plane of incidence.
For $s$-polarization there is no longitudinal wave and the second term of the susceptibility (\ref{eq:susc}) does not contribute, effectively reducing the susceptibility tensor to the diagonal form $\delta_{ij}\chi_\perp({\bm k},\omega)$ used by Halevi and Fuchs \cite{Halevi}. The derivation in this case is identical to their work and will not be repeated here.
In contrast, the $p$-polarization includes the longitudinal wave and the second term in (\ref{eq:susc}) contributes. We proceed to analyze this case.

\subsection{Field Equations}

For $p$-polarized light $E_y=0$, $E_x\ne0$ and $E_z\ne0$.
After a Fourier transform in the $xy$-plane, we write the $x$ and $z$ components of the wave equation (\ref{eq:wave_equation}) inside the material ($z>0$). Using equation (\ref{eq:infinite_polarization}) and the ansatz (\ref{eq:E_ansatz}) these components are:
\begin{align}
\sum_{n=1}^3
\bigg\{&
\left[
k_0^2(1+\chi_{xx}(q_n))-q_n^2
\right]\tilde{E}_x^{(n)}
\nonumber\\&
+
\left[
Kq_n+k_0^2\chi_{xz}(q_n)
\right]\tilde{E}_z^{(n)}
\bigg\}
e^{iq_nz}
=0,
\end{align}
\begin{align}
\sum_{n=1}^3
\bigg\{&
\left[
Kq_n+k_0^2\chi_{zx}(q_n)
\right]\tilde{E}_x^{(n)}
\nonumber\\&
+
\left[
k_0^2(1+\chi_{zz}(q_n))-K^2
\right]\tilde{E}_z^{(n)}
\bigg\}
e^{iq_nz}
=0.
\end{align}
These must hold for all values of $z$, giving:
\begin{align}
&
\left[
k_0^2(1+\chi_{xx}(q_n))-q_n^2
\right]\tilde{E}_x^{(n)}
\nonumber\\&+
\left[
Kq_n+k_0^2\chi_{xz}(q_n)
\right]\tilde{E}_z^{(n)}
=0,
\label{eq:Ex}
\end{align}
\begin{align}
&
\left[
Kq_n+k_0^2\chi_{zx}(q_n)
\right]\tilde{E}_x^{(n)}
\nonumber\\&+
\left[
k_0^2(1+\chi_{zz}(q_n))-K^2
\right]\tilde{E}_z^{(n)}
=0.
\label{eq:Ez}
\end{align}
If $\tilde{E}_x^{(n)}$ and $\tilde{E}_z^{(n)}$ are non-zero, the determinant of these equations must vanish for all three values of $n$. This requirement leads to the  dispersion relation (\ref{eq:transverse_disp_rel}) for $n=1,2$ and (\ref{eq:longitudinal_disp_rel}) for $n=3$. For the form of $\chi_\perp$ and $\chi_\parallel$ used here these dispersion relations are:
\begin{gather}
\left[
\left(1+\chi_0\right)k_0^2
-K^2
-
q_n^2
\right]
\left[
\Gamma_\perp^2
-
q_n^2
\right]
=
k_0^2
\frac{\omega_p^2}{\sigma_\perp^2},
\label{eq:q1q2}   \\
\left[
\Gamma_\parallel^2
-
q_3^2
\right]
=
\frac{\omega_p^2}{\left(1+\chi_0\right) \sigma_\parallel^2}.
\label{eq:q3}
\end{gather}
Rearranging (\ref{eq:Ex}) and (\ref{eq:Ez}) gives relations between the components of ${\bm \tilde{E}}$ :
\begin{align}
\tilde{E}_z^{(n)}=\eta^{(n)}\tilde{E}_x^{(n)},
\label{eq:Ex_Ez}
\end{align}
where $\eta^{(n)}$ takes the role of $\gamma^{(n)}$ in the Halevi \& Fuchs derivation\cite{Halevi}, and $\eta^{(1)}=-K/q_1$, $\eta^{(2)}=-K/q_2$ and $\eta^{(3)}=q_3/K$.

\subsection{Surface Impedance}
The reflection coefficient will be calculated below from the surface impedance, which for $p$-polarized light is given by:
\begin{align}
Z_p
=
\frac{
E_x(0^+)
}{
H_y(0^+)
}.
\end{align}
(Here ${\bm H}=\mu_0{\bm B}$.) The magnetic field $B_y$ can be expressed in terms of the electric field using $k_0{\bm B}={\bm k}\times{\bm E}$ and (\ref{eq:E_ansatz}): 
\begin{align}
B_y^{(n)}&
=
\frac{1}{k_0}
\left[
q_nE_x^{(n)}
-
KE_z^{(n)}
\right]
e^{iq_nz}
\nonumber\\&
=
\left[
\frac{
q_n
-
K\eta^{(n)}
}{k_0}
\right]
E_x^{(n)}
e^{iq_nz}
\nonumber\\&
=
\tau^{(n)}
E_x^{(n)}
e^{iq_nz}.
\end{align}
Here we have substituted for $E_z$ using (\ref{eq:Ex_Ez}) and defined $\tau^{(n)}$ by:
\begin{align}
\tau^{(n)}=
\frac{q_n^2+K^2}{q_n k_0}
=
\frac{k_0}{q_n}
\left(
1
+
\chi_0
+
\frac{\omega_p^2/\sigma_\perp^2}{q_n^2-\Gamma_\perp^2}
\right)
\end{align}
for $n=1,2$ and $\tau^{(3)}=0$ for the longitudinal wave.
The surface impedance can now be expressed in terms of field amplitude ratios:
\begin{align}
Z_p
=&
\frac{1}{\mu_0}
\frac{
E_x^{(1)}
+
E_x^{(2)}
+
E_x^{(3)}
}{
\tau^{(1)}
E_x^{(1)}
+
\tau^{(2)}
E_x^{(2)}
}
=
\frac{1}{\mu_0}
\frac{1
+
\frac{E_x^{(2)}}{E_x^{(1)}}
+
\frac{E_x^{(3)}}{E_x^{(1)}}
}{
\frac{q_1^2+K^2}{q_1 k_0}
+
\frac{q_2^2+K^2}{q_2 k_0}
\frac{E_x^{(2)}}{E_x^{(1)}}
}.
\label{eq:zp_ratios}
\end{align}

\subsection{Additional Boundary Conditions}
At this point we require the field amplitude ratios of the transmitted waves to find the surface impedance (\ref{eq:zp_ratios}).
By using the relation in (\ref{eq:Ex_Ez}), we rewrite the additional boundary conditions in (\ref{eq:abc}) solely in terms of $E_x$:
\begin{align}
\sum_n
\sum_j
\phi_{ij}^{(n)}
E^{(n)}_j&
=
\sum_n
\left[
\phi_{ix}^{(n)}
E^{(n)}_x
+
\phi_{iz}^{(n)}
E^{(n)}_z
\right]
\nonumber\\&
=
\sum_n
\left[
\phi_{ix}^{(n)}
+
\phi_{iz}^{(n)}
\eta^{(n)}
\right]
E^{(n)}_x
=
0.
\end{align}
We collect together the terms in square brackets to new variables $a_n$ and $b_n$ for $i=x$ and $z$ respectively:
\begin{align}
\sum_n
\left[
\phi_{xx}^{(n)}
+
\phi_{xz}^{(n)}
\eta^{(n)}
\right]
E^{(n)}_x
=&
\sum_n
a_n
E^{(n)}_x
=
0,
\nonumber\\
\sum_n
\left[
\phi_{zx}^{(n)}
+
\phi_{zz}^{(n)}
\eta^{(n)}
\right]
E^{(n)}_x
=&
\sum_n
b_n
E^{(n)}_x
=
0.
\label{eq:abc1}
\end{align}
The $\psi$ terms in (\ref{eq:abc}) are collected in a similar fashion to define $c_n$ and $d_n$:
\begin{align}
\sum_n
\left[
\psi_{xx}^{(n)}
+
\psi_{xz}^{(n)}
\eta^{(n)}
\right]
E^{(n)}_x
=&
\sum_n
c_n
E^{(n)}_x
=
0,
\nonumber\\
\sum_n
\left[
\psi_{zx}^{(n)}
+
\psi_{zz}^{(n)}
\eta^{(n)}
\right]
E^{(n)}_x
=&
\sum_n
d_n
E^{(n)}_x
=
0.
\label{eq:abc2}
\end{align}
We now have four ABC equations, compared to the two in the Halevi and Fuchs derivation\cite{Halevi}, which must all be satisfied.
With some manipulation we obtain:
\begin{align}
\frac{E_x^{(2)}}{E_x^{(1)}}
=\frac{
(3,1)_{\mu\nu}
}{
(2,3)_{\mu\nu}
},
\qquad
\frac{E_x^{(3)}}{E_x^{(1)}}
=\frac{
(1,2)_{\mu\nu}
}{
(2,3)_{\mu\nu}
},
\label{eq:amp_ratios}
\end{align}
where we define the symbol $(i,j)_{\mu\nu}=\mu_i\nu_j-\mu_j\nu_i$ with $\mu,\nu \in \{a,b,c,d\}$ and $\mu\neq\nu$. The field amplitude ratios in (\ref{eq:amp_ratios}) must give the same value for any combination of $\mu$ and $\nu$ ($\mu\neq\nu$).
Given the fact that $a$ and $c$ contain only $U_{xx}$ and $U_{xz}$ while $b$ and $d$ contain only $U_{zx}$ and $U_{zz}$, there must be some restrictions on the values that $U_{ij}$ can take. We find that (\ref{eq:amp_ratios}) can be satisfied for all $\mu,\nu$ combinations with $U_{xx}=U_{zx},U_{xz}=U_{zz}$, so that $b_n=(-k/\Gamma_\perp)a_n$ and $d_n=(\Gamma_\parallel/k)c_n$, reducing (\ref{eq:abc1}) and (\ref{eq:abc2}) to two equations.
Under these conditions we can make clear comparisons to the choice of ABC's presented by Halevi and Fuchs\cite{Halevi}, by associating  their $U_x$ and $U_z$ with $U_{xx}$ and $U_{zz}$ as in Table \ref{tab:abc}.

The choice of ABC is typically dependent on the type of material, with various authors making arguments based on the microscopic behavior of the system.
Both the Pekar\cite{pekar1958a,pekar1958b,pekar1958c,pekar1959} and 
Rimbey-Mahan\cite{rimbey1974,rimbey1975,rimbey1976,rimbey1977,rimbey1978} ABC were developed for Frenkel (tight-binding) excitons systems such as molecular crystals, although the second excluded the coupling of light to longitudinal modes of the medium. 
Ting \emph{et al.}\cite{ting1975} looked at a crystal model with Wannier-Mott (weak-binding) excitons, typically found in semiconductors.
The Fuchs-Kliewer\cite{ting1975,kliewer1968,kliewer1971,kliewer1975,ruppin1981} ABC considered a metal with specular reflection of electrons at the inner surface.
The Agarwal \emph{et al.}\cite{agarwal1971a,agarwal1971b,agarwal1972,birman1972,agarwal1973,maradudin1973,birman1974,mills1974,foley1975,bishop1976}
 ABC was not for a specific type of material, but derived under the assumption that changes in the susceptibility arising from the presence of the boundary can be neglected when considering bulk effects such as reflection and refraction. Henneberger\cite{hen98} considered a thin surface layer on the boundary as a source of radiation and found the ABC of Ting \emph{et al.}\cite{ting1975} in a simple case.

\begin{table}[!htb]\centering
\caption{\label{tab:abc}List of ABC's}
\begin{ruledtabular}
\begin{tabular}{llll}
 & $U_{xx}$ & $U_{yy}$ & $U_{zz}$\\
 \hline
Agarwal \emph{et al}.\cite{agarwal1971a,agarwal1971b,agarwal1972,birman1972,agarwal1973,maradudin1973,birman1974,mills1974,foley1975,bishop1976}
 & \phantom{-}0 & \phantom{-}0 & \phantom{-}0\\
Ting \emph{et al}.\cite{ting1975}& \phantom{-}1 & \phantom{-}1 & \phantom{-}1\\
Fuchs-Kliewer\cite{ting1975,kliewer1968,kliewer1971,kliewer1975,ruppin1981} & \phantom{-}1 & \phantom{-}1 & -1\\
Rimbey-Mahan\cite{rimbey1974,rimbey1975,rimbey1976,rimbey1977,rimbey1978} & -1 & -1 & \phantom{-}1\\
Pekar\cite{pekar1958a,pekar1958b,pekar1958c,pekar1959} & -1 & -1 & -1\\
\end{tabular}
\end{ruledtabular}
\end{table}

Using (\ref{eq:amp_ratios}), the surface impedance (\ref{eq:zp_ratios}) can be written:
\begin{align}
Z_p
=
\frac{1}{\mu_0}
\frac{
(2,3)_{ac}
+
(3,1)_{ac}
+
(1,2)_{ac}
}{
\frac{k_1^2}{q_1 k_0}
(2,3)_{ac}
+
\frac{k_2^2}{q_2 k_0}
(3,1)_{ac}
},
\label{eq:surface_imp}
\end{align}
where $k_n^2=K^2+q_n^2$. From now on we will only use the combination $\mu=a,\nu=c$ in the symbol $(i,j)_{\mu\nu}$, as in (\ref{eq:surface_imp}), so hereafter we omit the subscript $ac$ for notational simplicity.
In the simplest ABC case $U_{ij}=0$ and we find:
\begin{widetext}
\begin{align}
Z_p
=
-
\frac{k_0}{\mu_0}
\Bigg\{
\Gamma_\perp
-
\frac{
\left[
K^4
+K^2(q_1^2+q_1q_2+q_2^2)
+(q_1+q_2)q_1q_2q_3
\right]
-
\frac{\left(\Gamma_\parallel-\Gamma_\perp\right)(K^2+q_3^2)}{\left(\Gamma_\perp^2+K^2\right)+\left(\Gamma_\parallel-\Gamma_\perp\right)\left(\Gamma_\perp+q_3\right)}
\left[
q_1q_2(q_1+q_2)
\right]
}{
\left[
K^2(q_1+q_2-q_3)
+q_1q_2q_3
\right]
+
\frac{\left(\Gamma_\parallel-\Gamma_\perp\right)(K^2+q_3^2)}{\left(\Gamma_\perp^2+K^2\right)+\left(\Gamma_\parallel-\Gamma_\perp\right)\left(\Gamma_\perp+q_3\right)}
\left[
K^2-q_1q_2
\right]
}
\Bigg\}^{-1}
.
\end{align}
\end{widetext}
In the general case where $U_{xx}=U_{zx}$ and $U_{xz}=U_{zz}$, (\ref{eq:surface_imp}) reduces to the Halevi and  Fuchs result in the $\delta\to0$ limit where  $\chi_{ij}=\delta_{ij}\chi_\perp$.

\subsection{Reflection Coefficient}
\label{sec:reflection}
Using the vacuum surface impedance $Z_p^{(0)}=\sqrt{k_0^2-K^2}/\mu_0k_0$ and (\ref{eq:surface_imp}), we can construct the $p$-polarization reflection coefficient\cite{kliewer1968}:
\begin{align}
r_p=\frac{
Z_p^{(0)}-Z_p
}{
Z_p^{(0)}+Z_p
},
\label{eq:rp}
\end{align}
where:
\begin{align}
r_p
=\frac{{ E_r}}{{ E_0}}.
\label{eq:rp_definition}
\end{align}

\begin{figure}[!htb]\centering
{\includegraphics[width=\linewidth]{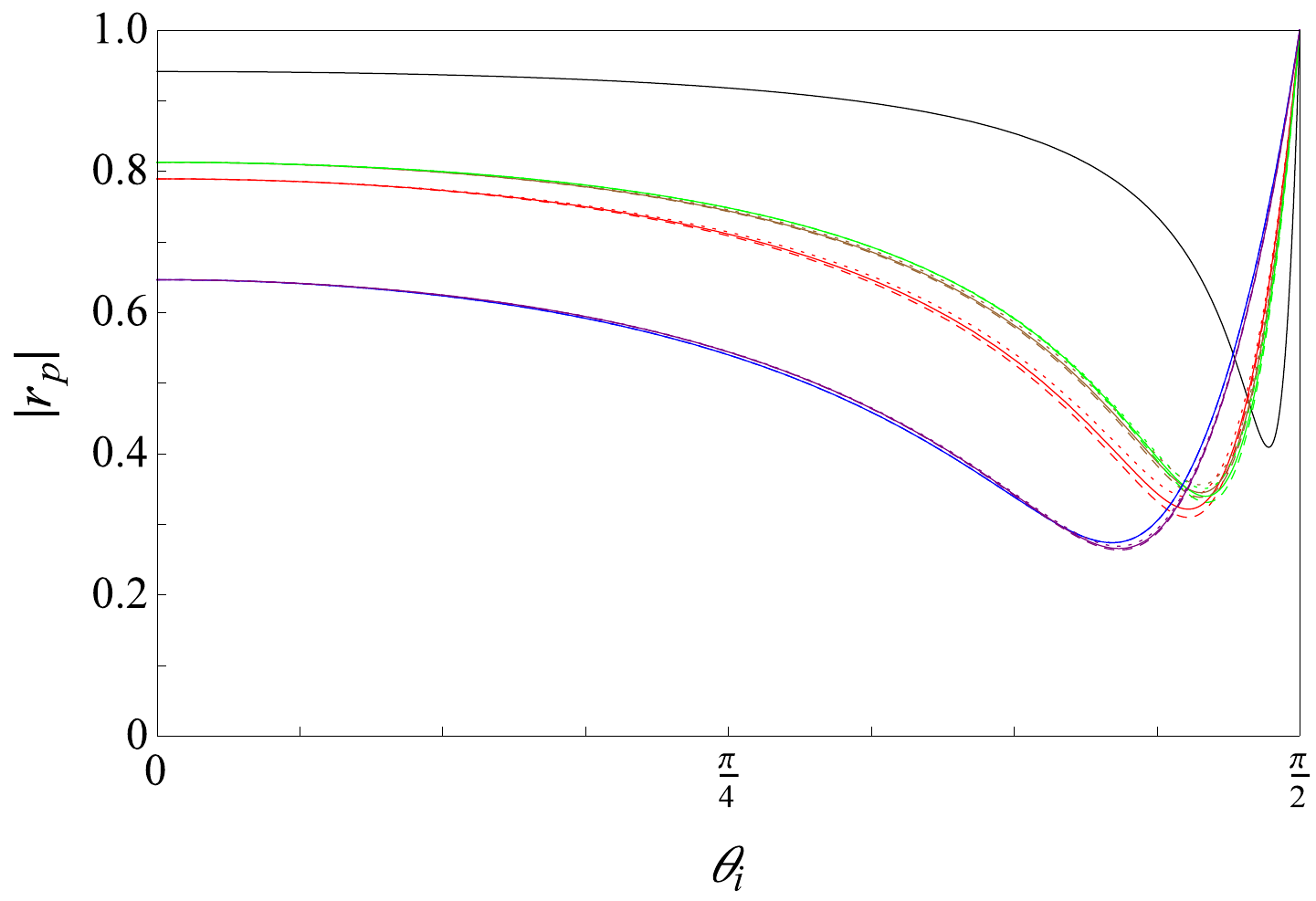}}
\caption{
Absolute value of the reflection coefficient of ZnSe at $\omega=\omega_T$ as a function of incident angle $\theta_i$ for propagating waves with $\delta=0$ (solid lines), $\delta=0.5$ (dashed) and $\delta=-0.5$ (dotted).
Includes Agarwal \emph{et al}. (Red), Ting \emph{et al}. (Brown), Fuchs-Kliewer (Green), Rimbey-Mahan (Blue) and Pekar (Purple) ABC's. The black curve has the spatial dispersion removed ($\sigma_\parallel=\sigma_\perp=0$).
}
\label{fig:rp_prop_angle}
\end{figure}

\begin{figure}[!htb]\centering
{\includegraphics[width=\linewidth]{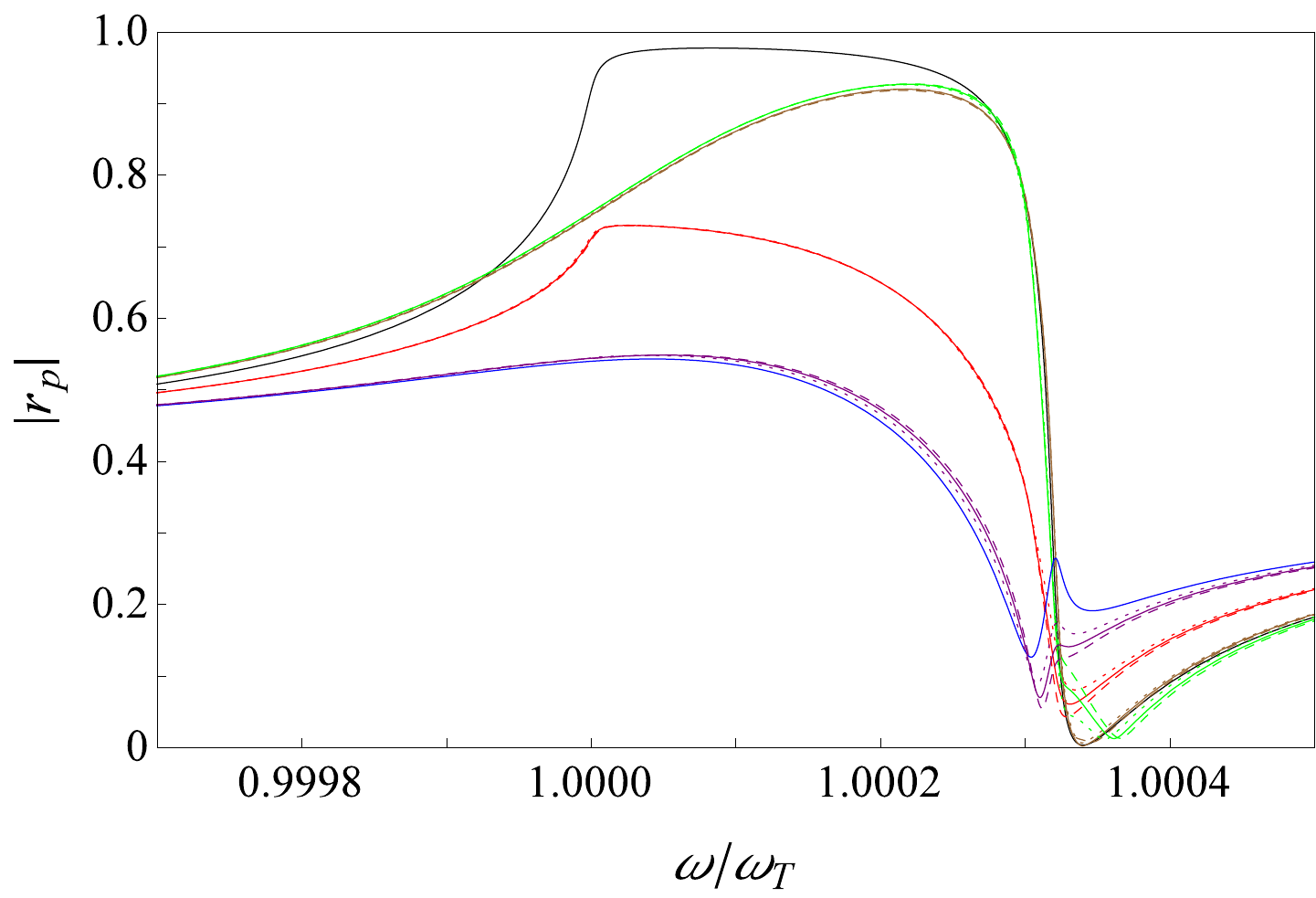}}
\caption{
Absolute value of the reflection coefficient of ZnSe at $\theta_i=\pi/4$ as a function of $\omega$.
Plot styles follow the conventions in Fig \ref{fig:rp_prop_angle}.
The parameter $\delta$ has the greatest effect near the reflection minimum.
}
\label{fig:rp_freq}
\end{figure}

\begin{figure}[!htb]\centering
{\includegraphics[width=\linewidth]{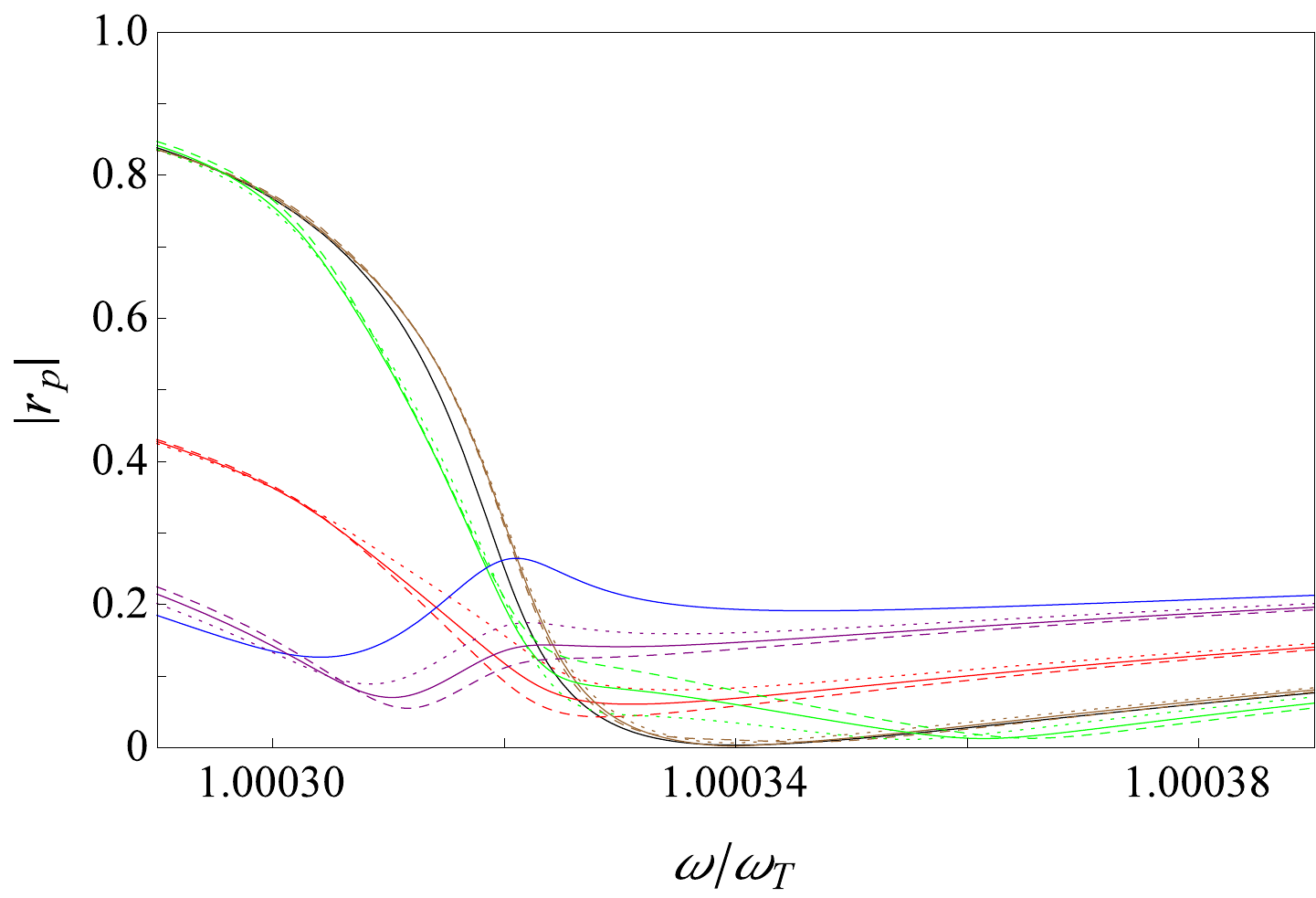}}
\caption{
Detail of Fig. \ref{fig:rp_freq} near the reflection minimum, where $\delta$ has the greatest effect.
}
\label{fig:rp_freq_detail}
\end{figure}

As an example, we consider parameters for ZnSe, the same material used by Halevi and Fuchs\cite{Halevi,ZnSe},
 with $\chi_0=8.1$, $\omega_p=3.25\times10^{14}$ rad s$^{-1}$, resonant frequency $\omega_T=4.25\times10^{15}$ rad s$^{-1}$ and damping $\gamma=4.25\times10^{10}$ rad s$^{-1}$. 
We define the non-local term $\sigma^2=\hbar\omega_T/(m_e+m_h)$ in the same manner as Halevi and Fuchs\cite{Halevi}, where $m_e$ and $m_h$ are the electron and hole mass.
For ZnSe, $\sigma_\perp=7.45\times10^5$m s$^{-1}$.
The exact value for $\delta$ is unknown, so we will present our results over the range $\delta=-0.5$ to $0.5$.

Figure \ref{fig:rp_prop_angle} shows the absolute value of the reflection coefficient for a range of ABC's and $\delta$ values at the resonant frequency $\omega_T$.
The choice of $U_{ij}$ values, specifically $U_{xx}$, has the greatest effect on $r_p$ near this frequency.
The $\delta$ parameter modifies the result to a much smaller extent, with the greatest change near the reflection minimum in the Agarwal \emph{et al}. ABC\cite{agarwal1971a,agarwal1971b,agarwal1972,birman1972,agarwal1973,maradudin1973,birman1974,mills1974,foley1975,bishop1976} , followed by Fuchs-Kleiwer\cite{ting1975,kliewer1968,kliewer1971,kliewer1975,ruppin1981}, Pekar\cite{pekar1958a,pekar1958b,pekar1958c,pekar1959} and Ting \emph{et al}\cite{ting1975}. The Rimbey-Mahan\cite{rimbey1974,rimbey1975,rimbey1976,rimbey1977,rimbey1978} result remains unchanged by $\delta$. This ABC was chosen so that no longitudinal wave could be generated, so in this case $\chi_\parallel$ and $\delta$ have no effect on $r_p$.

Figures \ref{fig:rp_freq} and \ref{fig:rp_freq_detail} show the $\omega$ dependence of $|r_p|$ for a fixed angle $\theta_i=\pi/4$.
It can be seen that $\delta$ has the greatest effect at frequencies slightly larger than $\omega_T$ near the reflection minima.
Agarwal \emph{et al}., Fuchs-Kliewer and Pekar are the most affected by $\delta$, while the change in Ting \emph{et al}. is significantly smaller.

\section{$p$-polarization transmission coefficients}
\label{sec:transmission}

We can find the transmission coefficients for the three transmitted waves by imposing the continuity of the tangential ${\bm E}$  field across the boundary.
Our choice of coordinate system means we simply equate the $E_x$ components on each side:
\begin{align}
\left[E_0-E_r\right]\cos{\theta_i}
=
\left[E^{(1)}_x+E^{(2)}_x+E^{(3)}_x\right].
\end{align}
By using (\ref{eq:amp_ratios}), (\ref{eq:rp_definition}) and $\cos{\theta_i}=\sqrt{k_0^2-K^2}/k_0$, this can be rewritten in terms of a single wave amplitude on the right:
\begin{align}
\frac{\sqrt{k_0^2-K^2}}{k_0}\left[1-r_p\right]E_0
=
\frac{\left[(2,3)+(3,1)+(1,2)\right]}{(2,3)}E^{(1)}_x.
\end{align}
Similar expressions can be found for $n=2,3$.
By using:
\begin{align}
E^{(n)}
=
\sqrt{
\left[E^{(n)}_x\right]^2
+
\left[E^{(n)}_z\right]^2
}
\end{align}
and (\ref{eq:Ex_Ez}), we derive the three transmission coefficients:
\begin{align}
t^{(1)}_p
=&
(2,3)\frac{k_1}{q_1}
\frac{\sqrt{k_0^2-K^2}}{k_0}\frac{\left[1-r_p\right]}{\left[(2,3)+(3,1)+(1,2)\right]}
,
\nonumber\\
t^{(2)}_p
=&
(3,1)\frac{k_2}{q_2}
\frac{\sqrt{k_0^2-K^2}}{k_0}\frac{\left[1-r_p\right]}{\left[(2,3)+(3,1)+(1,2)\right]}
,
\nonumber\\
t^{(3)}_p
=&
(1,2)\frac{k_3}{K}
\frac{\sqrt{k_0^2-K^2}}{k_0}\frac{\left[1-r_p\right]}{\left[(2,3)+(3,1)+(1,2)\right]}
,
\label{eq:tp}
\end{align}
where:
\begin{align}
t^{(n)}_p
= \frac{{ E^{(n)}}}{{E_0}}.
\label{eq:tp_definition}
\end{align}

\begin{figure}[!htb]\centering
{\includegraphics[width=\linewidth]{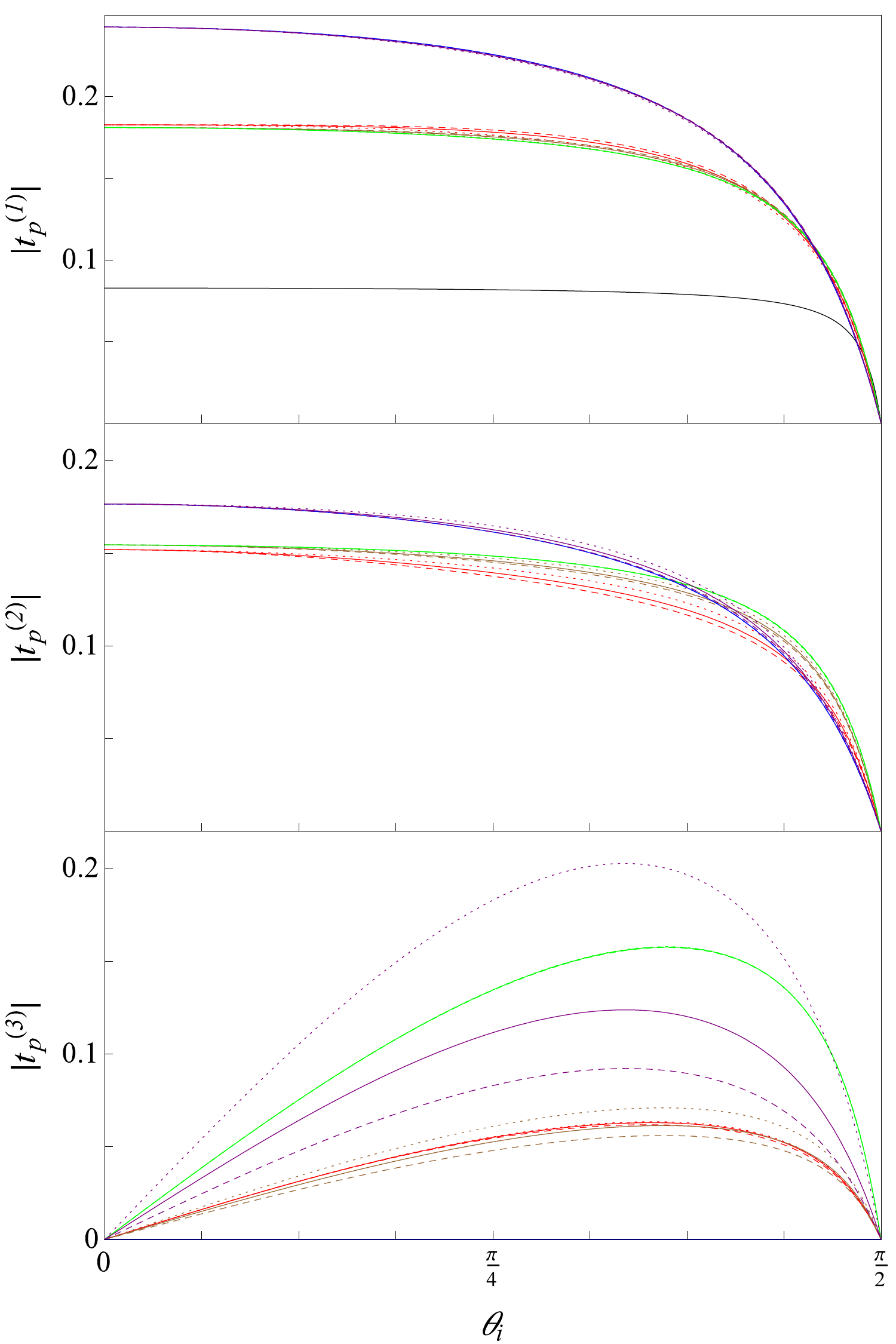}}
\caption{
Absolute values of the transmission coefficient for the three waves in ZnSe at $\omega=\omega_T$ as a function of incidence angle $\theta_i$.
Plot styles follow the conventions in Fig \ref{fig:rp_prop_angle}.
}
\label{fig:tp_prop_angle}
\end{figure}

Figure \ref{fig:tp_prop_angle} shows the absolute value of the transmission coefficients at $\omega=\omega_T$ as a function of incident angle for ZnSe.
For the two transverse waves, the $U_{ij}$ values, specifically $U_{xx}$, have the greatest effect.
Pekar and Rimbey-Mahan give near-identical results while Ting \emph{et al}. and Fuchs-Kleiwer also have similar values.
The effect of $\delta$ is negligible for $n=1$, which corresponds to the wave present when the material has local response, but its effect is larger for the second transverse wave introduced by the nonlocal dependence.
The longitudinal wave shows different behavior. 
At normal incidence $t_p^{(3)}=0$, since  ${\bm E_0}$ is polarized parallel to the surface and there is no $z$-component to excite the longitudinal wave.
The  results show a large spread with $\delta$. Fuchs-Kleiwer is generally the largest, while Rimbey-Mahan is always zero due to the absence of a longitudinal wave in that case.
Beyond these features, the behavior of $t_p^{(3)}$ for the various ABC's, and also the effect of $\delta$, is strongly dependent on the material parameters and frequency.
For example, while Agarwal \emph{et al.} and Ting \emph{et al.} give similar results in Fig. \ref{fig:tp_prop_angle} while Fuchs-Kleiwer is not affected by $\delta$, this is not true in general.

We have already highlighted the Rimbey-Mahan ABC, where the values of $U_{ij}$ leads to $c_1=c_2=0$. As a result $(1,2)=0$ and there is no longitudinal wave ($E^{(3)}=0$). We now check the possibility of choosing $U_{ij}$ to give no \emph{transverse} waves in the medium. With no transverse waves the ${\bm B}$-field in the medium is zero, leading to $Z_p=\infty$ and perfect reflection with $r_p=-1$.
This requires $a_3=0$ and $c_3=0$, leading to the following values of $U_{xx}$ and $U_{zz}$:
\begin{align}
U_{xx}
=
1+
\frac{
2q_3(\Gamma_\parallel^2+K^2)
}{
(\Gamma_\perp\Gamma_\parallel+K^2)(\Gamma_\parallel-q_3)
}
\nonumber\\
U_{zz}
=
-1+
\frac{
2\Gamma_\perp(\Gamma_\parallel^2+K^2)
}{
(\Gamma_\perp\Gamma_\parallel+K^2)(\Gamma_\parallel-q_3)
}.
\label{eq:long_only_1}
\end{align}
In the $\delta=0$ case, this reduces to:
\begin{align}
U_{zz}
=
U_{xx}
=
\frac{\Gamma+q_3}{\Gamma-q_3}.
\label{eq:long_only_2}
\end{align}
Using the $q_3$ definition in (\ref{eq:q3}) we find from (\ref{eq:long_only_2}) the required $U$ to get perfect reflection for given $K$ and $\omega$ values in the $\delta=0$ case:
\begin{align}
U_{xx}
=U_{zz}
=
\frac{
1+\sqrt{
1+\frac{1}{1+\chi_0}\frac{\omega_p^2}{\omega_T^2+\sigma^2K^2-\omega^2-i\gamma\omega}
}
}{
1-\sqrt{
1+\frac{1}{1+\chi_0}\frac{\omega_p^2}{\omega_T^2+\sigma^2K^2-\omega^2-i\gamma\omega}
}
}.
\label{eq:longU}
\end{align}
The relation gives $\left|U_{xx}\right|>1$, conflicting with the definition of $U$ as the reflection coefficient of the polarization waves at the surface. Note that (\ref{eq:longU}) does not allow $\left|U_{xx}\right|=1$ because the square-root quantity does not vanish for any real $K$ and $\omega$. Thus we must have transverse transmitted waves in the material.

\section{Spectral Energy density}\label{sec:energy}

\begin{figure*}[!htb]
{\includegraphics[width=178mm]{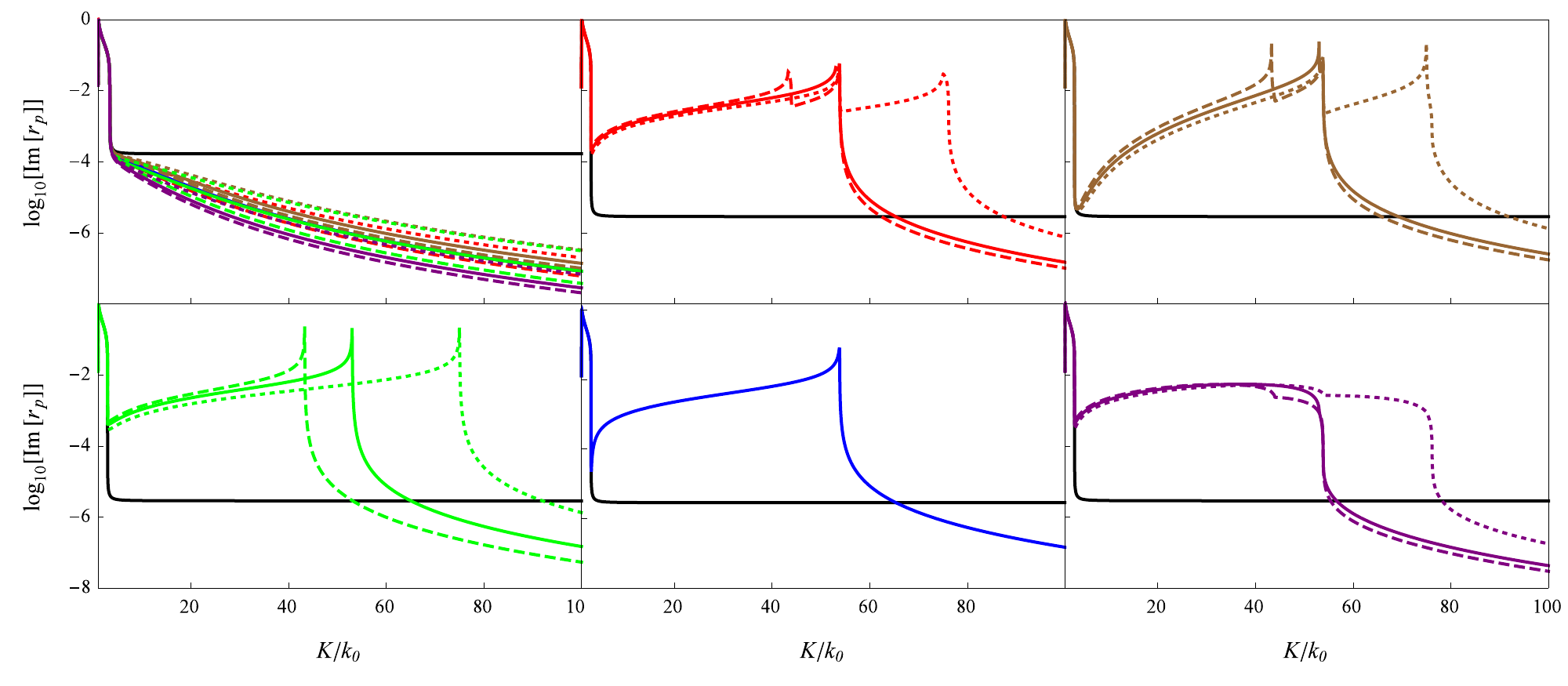}}
\caption{
Logarithmic plot of Im$[r_p]$ used in the $u_{tot}$ integration (\ref{eq:utot}) at $\omega=0.999\omega_T$ (top left) and $\omega=1.01\omega_T$ (all others) as a function of $K$ for evanescent waves.
Plot styles follow the conventions in Fig \ref{fig:rp_prop_angle}.
}
\label{fig:rp_evan_k}
\end{figure*}

We now apply the previous results to the problem of electromagnetic zero-point and thermal radiation near material boundaries. It is well known that the neglect of spatial dispersion leads to an unphysical divergence in the energy density of thermal radiation at a planar boundary\cite{henkel2000,joulain2005}. The divergence occurs at the level of the spectral  energy density (i.e.\ the energy density per unit frequency) and this same divergence is present for the zero-point spectral  energy density. Although the total zero-point energy density will always diverge if it is not regularized, the \emph{spectral} energy density of zero-point radiation should be finite without regularization\cite{hor14}. When the nonlocal response of materials is taken into account, all these spurious divergences must disappear and thus spatial dispersion is the key property that determines the spectral energy density of zero-point and thermal radiation near material boundaries. It has already been shown\cite{henkel2006} that a plasma described by the nonlocal Lindhard susceptibility gives a finite spectral energy density at a planar boundary. A similar model removes an unphysical divergence in spectral zero-point and thermal correlations inside a homogeneous material\cite{nar10,hor14}. Here we show that the quite general dielectric model used here is free of the aforementioned divergences at a planar boundary, and we also show that differences between the transverse and longitudinal susceptibilities can have a large effect.

The average energy density of zero-point and thermal radiation in the vacuum region outside the semi-infinite dielectric of Fig.~\ref{fig:model} is given by:
\begin{align}
\langle U\rangle
=&
\frac{\varepsilon_0}{2}\langle \left|{\bm E}\left({\bm r},t\right) \right|^2\rangle
+
\frac{\mu_0}{2}\langle \left|{\bm B}\left({\bm r},t\right) \right|^2\rangle
\nonumber\\
=&
\int_0^\infty
d\omega \, u_{tot}\left(z,\omega\right),
\end{align}
where $u_{tot}(z,\omega)$ is the spectral energy density that depends on $z$. We assume that the semi-infinite dielectric is in thermal equilibrium with the surroundings and we will include the zero-point contribution. The  expression for $u_{tot}(z,\omega)$ can be written in terms of the reflection coefficients for $s$- and $p$-polarized light at the planar boundary (see for example Ref.~\onlinecite{kit05}):
\begin{align}
 & \!\!\! u_{tot}(z,\omega)
=
\nonumber\\
 &  \!\!\! \frac{u_0}{k_0}
\int_0^{k_0}  \!\!\!
\frac{KdK}{\sqrt{k_0^2-K^2}}
\left[
1+\frac{K^2\textrm{Re} \left[ (r_s+r_p) e^{-2i\sqrt{K^2-k_0^2}z}  \right] }{2k_0^2}
\right]
\nonumber\\
& +
\frac{u_0}{2 k_0^3} \int_{k_0}^\infty
\frac{K^3dK}{\sqrt{K^2-k_0^2}}
\textrm{Im}[r_s + r_p]
e^{2\sqrt{K^2-k_0^2}z}
,
\label{eq:utot}
\end{align}
where $u_0$ is the spectral energy density in the absence of the material, given by:
\begin{align}
u_0&
=
\frac{\Theta(\omega,T)\omega^2}{\pi^2c^3},  \\
\Theta(\omega,T)&
=
\hbar\omega
\left(
\frac{1}{2}
+\frac{1}{e^{\hbar\omega/k_BT}-1}
\right).
\end{align}
The quantity $\Theta(\omega,T)$ is the mean energy of a harmonic oscillator in thermal equilibrium, the first term of which gives rise to the electromagnetic zero-point energy. The first term in (\ref{eq:utot}) is the contribution of propagating waves while the second term comes from evanescent waves.

If spatial dispersion is ignored then as $K\to\infty$ the reflection coefficients have the limits $r_s\to0$, $r_p\to\chi(\omega)/(2+\chi(\omega))$, where $\chi(\omega)$ is the local susceptibility of the isotropic medium.
For large $K$ the $r_p$-term in the second integral in (\ref{eq:utot}) is then proportional to $K^2$, dominating the final result for $u_{tot}(z,\omega)$ when $z$ is much smaller than the wavelength. This leads to a simple approximate expression\cite{joulain2005} for the second integral in (\ref{eq:utot}) as $z\to0$:
\begin{align}
\frac{1}{4z^3}
\frac{\textrm{Im}[\chi(\omega)]}{|2+\chi(\omega)|^2},
\label{eq:utot_nondispersive}
\end{align}
which diverges as $z\to0$ for complex $\chi(\omega)$. This unphysical divergence is removed when the bulk susceptibility has a dependence on $k$ of the form of the second term in (\ref{eq:susceptibility_basic}), which is the form we used for the transverse and longitudinal susceptibilities. But the background term $\chi_0$ in (\ref{eq:susceptibility_basic}) will still lead to a divergence in the spectral energy density if it is complex. As already noted, the background term $\chi_0$ should be replaced by additional resonance terms in a more general susceptibility $\chi(k,\omega)$. For our parameters for ZnSe, however, $\chi_0$ is real and so it causes no difficulties in the numerical calculations below.

We now consider the spectral energy density (\ref{eq:utot}) for our model with bulk tensor susceptibility (\ref{eq:susc}). The $p$-polarization reflection coefficient $r_p$ was found in Sec.~\ref{sec:p-polarization} (see (\ref{eq:rp})) and $r_s$ is given by Halevi and Fuchs\cite{Halevi}:
\begin{align}
r_s=\frac{
Z_s^{(0)}-Z_s
}{
Z_s^{(0)}+Z_s
},
\label{eq:rs}
\end{align}
where $Z_s^{(0)}=k_0/\mu_0\sqrt{k_0^2-K^2}$ and:
\begin{widetext}
\begin{align}
Z_s=
\frac{(1+U_{yy})k_0(q_1q_2+\Gamma_\perp^2)
+(1-U_{yy})k_0\Gamma_\perp(q_1+q_2)
}{(1+U_{yy})q_1q_2(q_1+q_2)
+(1-U_{yy})\Gamma_\perp(q_1^2+q_1q_2+q_2^2-\Gamma_\perp^2)}.
\label{eq:Zs}
\end{align}
\end{widetext}
We will now substitute these refection coefficients into (\ref{eq:utot}) and perform the integration over $K$, with the same material parameters as used previously. As we have seen, it is the behavior of the $r_p$ term in the second (evanescent wave) integral in (\ref{eq:utot}) that determines the spectral energy density near the boundary.

Figure \ref{fig:rp_evan_k} shows the behavior of $\textrm{Im}[r_p]$ at $\omega=0.999\omega_T$ and $\omega=1.01\omega_T$ for evanescent waves ($K>k_0$). In contrast to the $r_p$ results for propagating waves in Sec \ref{sec:p-polarization}, $\delta$ has a significant effect on $\textrm{Im}[r_p]$ for evanescent waves.
For $K<\sqrt{(1+\chi_0)k_0^2}$, the reflection coefficient closely matches the local result, whereas for $K>\sqrt{(1+\chi_0)k_0^2}$ the plots show how $\textrm{Im}[r_p]$ has a very different behavior from the local model. Spatial dispersion causes $\textrm{Im}[r_p]$ to fall off as $1/K^4$ for large $K$, but its behavior for smaller $K$ differs significantly for $\omega<\omega_T$ compared to $\omega>\omega_T$. For $\omega=0.999\omega_T$, Agarwal \emph{et al}., Fuchs-Kleiwer and Rimbey-Mahan are nearly identical in the $\delta=0$ limit, while Ting \emph{et al}. is larger and Pekar is smaller.
The $\delta$ parameter has the greatest effect on Fuchs-Kleiwer and a smaller effect on Agarwal \emph{et al}., Ting \emph{et al}. and Pekar. Rimbey-Mahan remains unchanged with $\delta$ due to the absence of the longitudinal wave. For $\omega=1.01\omega_T$ there is a peak in $\textrm{Im}[r_p]$, followed by a sharp drop, at the value of $K$ where $\textrm{Re}[\Gamma_\perp^2]$ changes sign to a negative value, with a similar peak at the value of $K$ where $\textrm{Re}[\Gamma_\parallel^2]$ changes sign. For $\delta=0$ we have $\Gamma_\perp=\Gamma_\parallel$ and there is only one such peak. The exceptions to this behavior are Rimbey-Mahan, which always displays only the $\Gamma_\perp$ peak and Fuchs-Kleiwer, which displays only the $\Gamma_\parallel$ peak. 

The large $K$ behavior of $\textrm{Im}[r_p]$ at all frequencies means the function in the evanescent integral of (\ref{eq:utot}) without the exponential is proportional to $1/K^2$ as $K\to\infty$. As a result, the integral over evanescent waves converges to a finite value even in the $z=0$ case.

\begin{figure}[!htb]\centering
{\includegraphics[width=\linewidth]{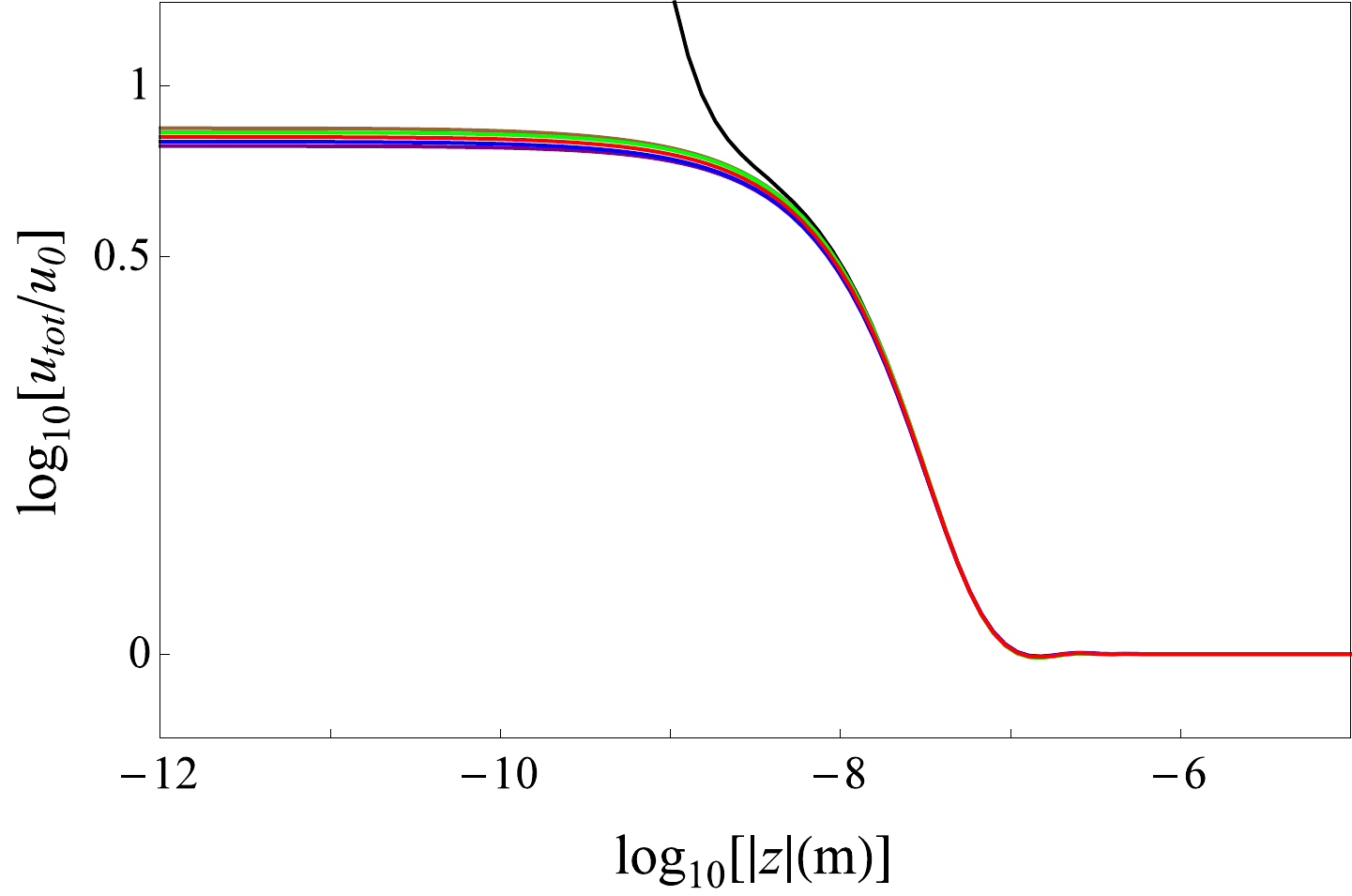}}
\caption{
Rescaled spectral energy density at $\omega=0.999\omega_T$ as a function of distance from the surface $z$.
Plot styles follow the conventions in Fig \ref{fig:rp_prop_angle}.
}
\label{fig:utotsmall}
\end{figure}

\begin{figure}[!htb]\centering
{\includegraphics[width=\linewidth]{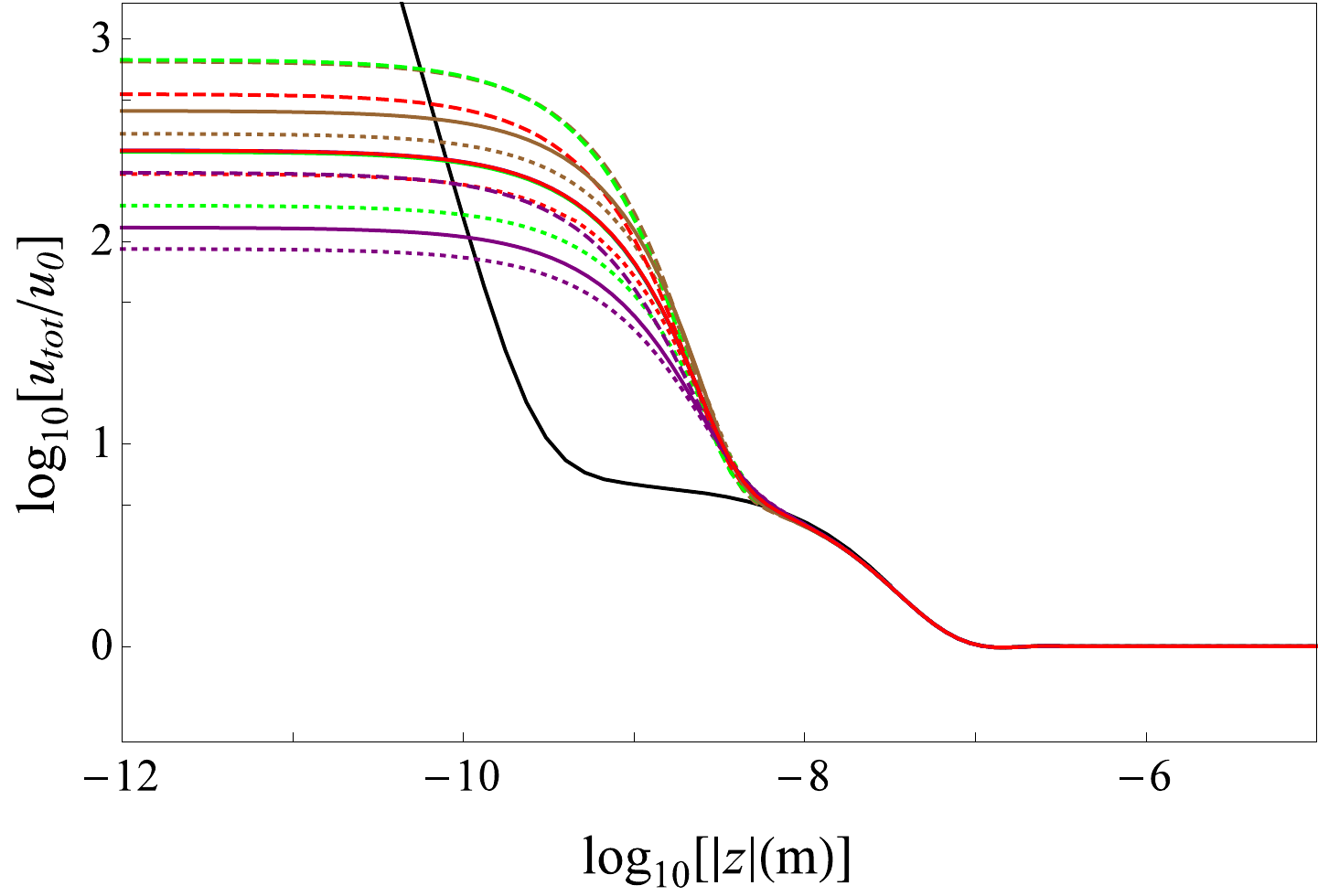}}
\caption{
Rescaled spectral energy density at $\omega=1.01\omega_T$ as a function of distance from the surface $z$.
Plot styles follow the conventions in Fig \ref{fig:rp_prop_angle}.
Agarwal \emph{et al}., Fuchs-Kleiwer and Rimbey-Mahan results are almost identical in the $\delta=0$ case.
}
\label{fig:utotlarge}
\end{figure}

Figures \ref{fig:utotsmall} and \ref{fig:utotlarge} show the spectral energy density $u_{tot}(z,\omega)$ divided by $u_0$  as function of distance from the boundary. Results for the various ABC's are shown together with the local result. Figure~\ref{fig:utotsmall} is for $\omega<\omega_T$ while figure~\ref{fig:utotlarge} is for $\omega>\omega_T$.
For the smaller frequency $\omega=0.999\omega_T$ (Fig. \ref{fig:utotsmall}), the integral is dominated by the smaller values of $K$  for which $\textrm{Im}[r_p]$ is very similar for all the ABC's. As a result the spectral energy density shows small differences between the ABC's while differing significantly from the local (diverging) result as $|z|\to0$. The effect of $\delta$ is negligible as it only affects large-$K$ values of $\textrm{Im}[r_p]$ that are already very small. The choice of $U_{ij}$ and $\delta$ is more significant at the larger frequency $\omega=1.01\omega_T$ (Fig. \ref{fig:utotlarge}).
In the $\delta\to0$ limit, the Agarwal \emph{et al}., Fuchs-Kleiwer and Rimbey-Mahan results are almost identical, while Ting \emph{et al}. is larger and Pekar is smaller.
Fuchs-Kleiwer is affected the most by $\delta$, followed by Ting \emph{et al}., Agarwal \emph{et al}. and finally Pekar, while Rimbey-Mahan remains unchanged.

Below 20nm for $0.999\omega_T$ and 8nm for $1.01\omega_T$, the spatially dispersive result begins to differ from the local medium.
These values of $|z|$ match the condition:
\begin{align}
\sigma^2\left(\frac{2\pi}{z}\right)^2=|\omega_T^2-\omega^2-i\gamma\omega|,  \label{scale1}
\end{align}
since this distance corresponds to the wavelength of the polarization waves. The nonlocal $u_{tot}$ begins to saturate to a finite value below 1nm, removing the divergent $1/z^3$ behavior of the local medium. This distance is given by:
\begin{align}
\sigma^2\left(\frac{2\pi}{z}\right)^2=\omega_T^2, \label{scale2}
\end{align}
corresponding to wavelengths of the polarization waves below which the nonlocal term starts to dominate the resonance term ($\omega_T^2$). Both of the distance scales (\ref{scale1}) and (\ref{scale2}) depend on the relevant material parameters. For very small distances, certainly below 1nm, this model is no longer valid as other effects need to be included, e.g.\  higher-order terms of $k$ in the denominator of $\chi$, surface roughness, and quantum properties of the surface.

\begin{figure}[!htb]\centering
{\includegraphics[width=\linewidth]{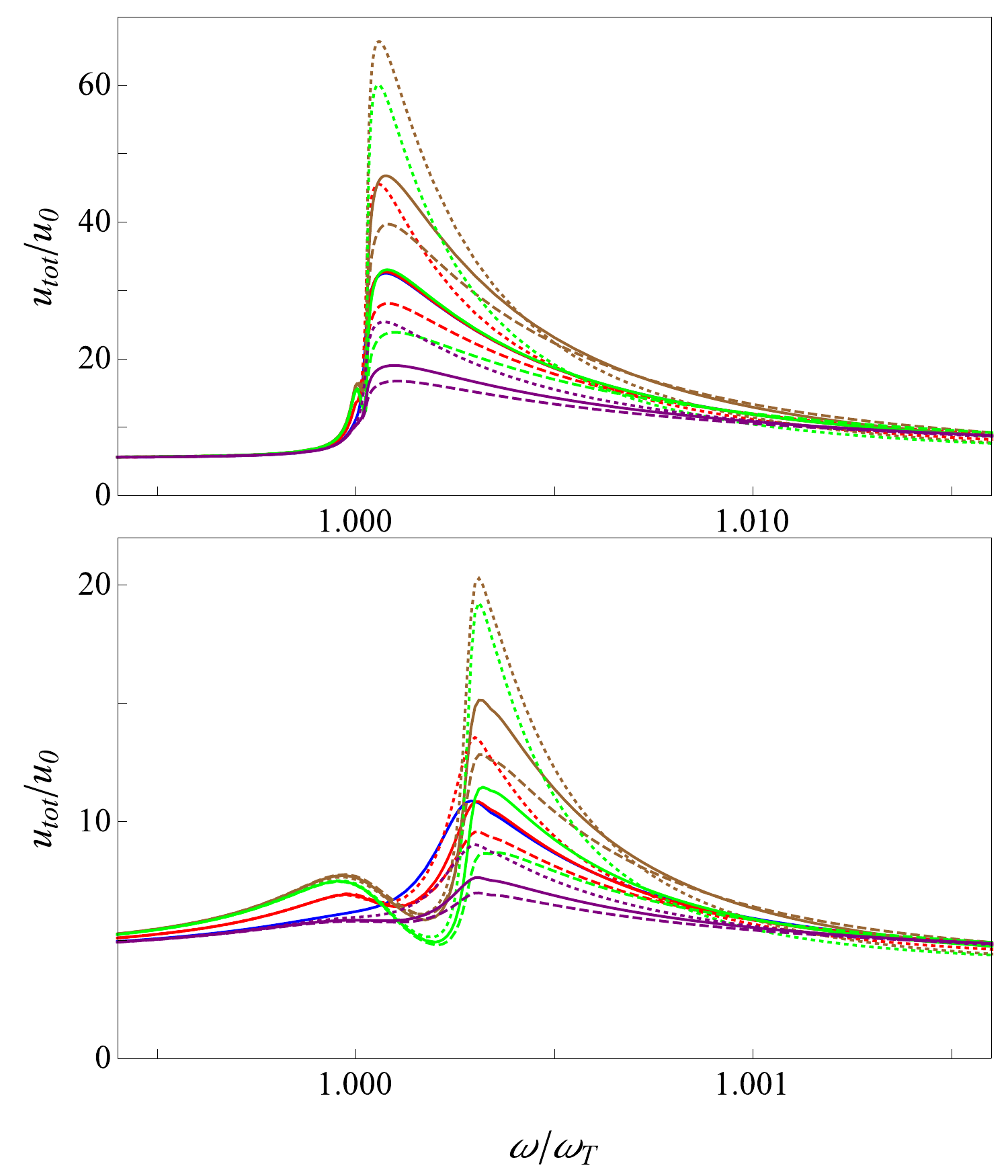}}
\caption{
Rescaled spectral energy density as a function of $\omega$ at a distance of 3nm (top) and 10nm (bottom) from the dielectric surface. Note the difference in scales.
Plot styles follow the conventions in Fig \ref{fig:rp_prop_angle}.
The Agarwal \emph{et al}., Fuchs-Kleiwer and Rimbey-Mahan results are almost identical in the $\delta=0$ case at $z$=3nm.
}
\label{fig:utotfreq}
\end{figure}

Figure \ref{fig:utotfreq} shows $u_{tot}$ as a function of frequency at fixed distances of 3nm and 10nm from the surface for the different ABC's.
Similar behavior can be observed in both cases, with two key features present.
The first is the small peak at $\omega_T$, which is a feature of the $\textrm{Im}[r_s]$ integral and as a result  is unaffected by $\delta$. The second, larger peak at higher frequencies occurs when the sign of $\textrm{Re}[\Gamma_\perp^2]$ and $\textrm{Re}[\Gamma_\parallel^2]$ can change with $K$ leading to peaks in Im$[r_p]$ as a function of $K$.
This larger peak in $u_{tot}$ increases and broadens as the surface is approached.
In the $\delta\to0$ limit,  Ting \emph{et al}. gives the largest value, Pekar is the smallest and 
Agarwal \emph{et al}., Fuchs-Kleiwer and Rimbey-Mahan all take very similar intermediate values. The peak is strongly dependent on the value of $\delta$;  for example, the greatest effect is in the Fuchs-Kleiwer peak, which varies by almost a factor of 3 over the range $-0.5<\delta<0.5$ at 3nm. 
The effect of $\delta$ decreases in the order Ting \emph{et al}., Agarwal \emph{et al}. and finally Pekar.
This strongly contrasts with the almost negligible effect $\delta$ has on the reflection coefficient for propagating waves.

From these results it is clear that the tensor nature of the susceptibility and the difference between $\chi_\perp$ and $\chi_\parallel$ must be taken into consideration when considering the spectral energy density of zero-point and thermal radiation close to material boundaries.

We note that the thermal energy density near metal surfaces has been probed using near-field microscopy\cite{wilde06}. The zero-point spectral energy density can be probed by measuring spontaneous emission rates close to a boundary\cite{pur46,dre68,bar98,Novotny}. In addition, curved boundaries experience a deforming force (Casimir ``self-force") due to the local zero-point and thermal radiation\cite{boy68,mil80,candelas1982,hor14}, although this effect will be difficult to measure experimentally in any direct manner. 

\section{Conclusion}\label{sec:conclusion}

We have derived exact expressions for reflection and transmission coefficients at a boundary of an isotropic spatially dispersive dielectric, taking into account that such a material has a tensor susceptibility.
Surface effects have been included by introducing phenomenological reflection coefficients $U_{ij}$ for polarization waves at the boundary. We have compared the effect of specific values of $U_{ij}$ corresponding to different ABC sets in the literature and also the effect of the inequality between the transverse and longitudinal susceptibilities ($\chi_\perp$ and $\chi_\parallel$ ). As noted by Halevi and Fuchs \cite{Halevi}, the coefficients $U_{ij}$ will in reality depend on frequency, in contrast to the simple constant values assumed in the ABC sets.

The reflection coefficient for $s$-polarization has already been derived by Halevi and Fuchs \cite{Halevi} so here we looked in detail at the  $p$-polarization reflection coefficient and transmission coefficients. For propagating waves, differences between  $\chi_\perp$ and $\chi_\parallel$ have the greatest effect on $r_p$ near the reflection minima, but it is the choice of ABC that has a far more significant effect on $r_p$. 

We also considered in detail the zero-point and thermal spectral energy density $u_{tot}(z,\omega)$ outside the dielectric. The inclusion of spatial dispersion naturally removes the $1/z^3$ divergence of the local-medium result so that $u_{tot}(z,\omega)$  attains a constant value at distances of the order of 1nm, depending on material parameters. The inequality between $\chi_\perp$ and $\chi_\parallel$ was found to have a very significant effect on the maximum of $u_{tot}(z,\omega)$ as a function of $\omega$, even at a distance of 10nm from the surface. These results demonstrate that divergences in the (regularized) zero-point energy density and stress at planar boundaries\cite{sop02,bar05} are due to the neglect of spatial dispersion\cite{hor14}. Similar divergences in zero-point and thermal radiation at curved boundaries should also be removed by including non-local response\cite{hor14} and this will enable proper estimates of Casimir self-forces on objects like the dielectric ball and spherical shell\cite{boy68,mil80,candelas1982}.

The ABCs investigated here arose from consideration of different materials and models (as described in Sec.~\ref{sec:p-polarization}). The question of which ABC is appropriate for a given dielectric is difficult to assess in practice, given the necessarily simplified analysis which leads to the ABCs. In the case of the conduction electrons of a metal, far more is known both from theory and experiment\cite{raz11,wie12,pen12,tos15,sch16}. Comparison of further experimental results with the predictions of different ABCs (such as those calculated here) may shed light on this interesting question.

\acknowledgments
We thank S.A.R. Horsley for helpful discussions.

\end{document}